\documentclass[12pt]{iopart}

\usepackage{amsfonts}
\usepackage{multirow}
\usepackage{graphicx}
\eqnobysec

\usepackage{color}
\definecolor{darkgreen}{rgb}{0,0.6,0}

\newcommand{\misc}[1]{}

\begin{document}



\title{Measurement and Ergodicity in Quantum Mechanics}

\author{Mariano Bauer and Pier A. Mello}
\address{Instituto de F\'{\i}sica, Universidad Nacional Aut\'{o}noma de M\'{e}xico,
M\'{e}xico, D.F. C.P. 04510, Mexico}

\begin{abstract}
The experimental realization of successive non-demolition measurements on single microscopic systems brings up the question of ergodicity in Quantum Mechanics (QM).
We investigate whether time averages over one realization of a single system are related to QM averages over an ensemble of similarly prepared systems. 
We adopt a generalization of von Neumann model of measurement, coupling the system to $N$ ``probes" 
--with a strength that is at our disposal-- 
and detecting the latter.
The model parallels the procedure followed in experiments on Quantum Electrodynamic cavities.
The modification of the probability of the observable eigenvalues due to the coupling to the probes can be computed analytically and the results compare qualitatively well with those obtained numerically by the experimental groups.
We find that the problem is not ergodic, except in the case of an eigenstate
of the observable being studied.
\end{abstract}

\date{today}



\section{Introduction}
\label{intro}

The question of ergodicity in Quantum Mechanics (QM) has long been studied, 
a ``quantum ergodic theorem" (QET) having been formulated by von Neumann in 1929 \cite{von_neumann_QET}.
Ref. \cite{goldstein_et_al_2010} discusses the investigations on the subject, from von Neumann's QET up to recent publications.
QM ergodicity for a {\em macroscopic} (more than $10^{20}$ particles) quantum system means \cite{goldstein_et_al_2010}
\begin{equation}
\lim_{T \to \infty} \frac{1}{T} 
\int_0^{\infty} 
\langle \psi(t)| \hat{A} | \psi(t)\rangle dt
= \frac{1}{d}\sum_{i\in d}\langle \psi_i |\hat{A}|\psi_i \rangle .
\label{ergodicity 1}
\end{equation}
The QM expectation values $\langle \psi(t)| \hat{A} | \psi(t)\rangle$ and 
$\langle \psi_i |\hat{A}|\psi_i \rangle$ are 
{\em averages over an ensemble of similarly prepared systems} \cite{ballentine}, 
to be called a {\em QM ensemble};
they result from measurements of the dynamical variable $\hat{A}$, independently of the invasive nature of the observations.
In addition, a {\em time average} appears on the LHS of
Eq. (\ref{ergodicity 1}) and, on the RHS, an {\em ensemble average} over the $d$ eigenstates of the microcanonical subspace.
As usual, the systems described by a microcanonical ensemble have fixed number of particles and volume, and an energy lying in an interval $\Delta \ll E$ containing $d$ levels \cite{pathria}.
However, Ref. \cite{goldstein_et_al_2010} remarks that the property described in the QET 
``is not precisely analogous to the standard notion of ergodicity as known from classical mechanics and the mathematical theory of dynamical systems".
Illustrations of the time average appearing in Eq. (\ref{ergodicity 1}) can be found in Ref. \cite{mello_mosh}.

Here we pose a question of a different kind, applicable to systems 
{\em not necessarily macroscopic}, 
motivated by theoretical considerations and experimental developments.
Non-demolition measurements distributed in time \cite{caves}
and ensembles of measurements on single microscopic systems
(few particles or field modes)
\cite{brune,guerlin_et_al07}
provide a situation closer to the classical notion of ergodicity \cite{yaglom}, as briefly 
described above and further explained after Eq. (\ref{bar_Q 1}), and 
in relation with Eqs. (\ref{ens average of t_av_1_syst 1 a}), 
(\ref{ens average of t_av_1_syst 1 b}) and 
(\ref{dispersion of t_av_1_syst 0})-(\ref{dispersion of t_av_1_syst f}).

\begin{figure}[h]
\centerline{
\includegraphics[scale=0.6]{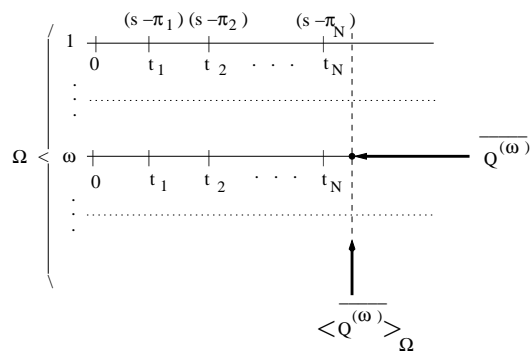}
}
\caption{
\footnotesize{Schematic representation of the setup used to discuss ergodicity:
the various realizations of the extended system are indicated by $\omega$.
Also shown are $\overline{Q^{(\omega)}}$ of Eq. (\ref{bar_Q 1})
and the QM ensemble average
$\left\langle \overline{Q^{(\omega)}} \right\rangle_{\Omega}$
of Eq. (\ref{ens average of t_av_1_syst 1 a}), both detected at the time $t_N^+$.
}
}
\label{fig1}
\end{figure}

The question will then be the following.
Out of a QM ensemble, consider {\em one single system} $s$;
at time $t_1$ we measure the observable $\hat{A}$ and again at $t_2$.
If the first measurement is very invasive, we disturb 
the system $s$ so much that the next measurement does not find 
$s$ in the original state.
We thus introduce the first stage of the measurement, or ``pre-measurement", explicitly in the QM description, by coupling $s$ to a ``probe" $\pi_1$ at time $t_{1}$;
we control the disturbance through the system-probe coupling strength. 
Next, we couple the {\em same} system $s$ to another probe $\pi_2$ at $t=t_2$ [see the model Hamiltonian of Eq. (\ref{V 2meas}) below, extended to $N$ probes], etc.;
each probe $\pi_i$, $i=1,\cdots, N$, interacts with the system proper at some instant $t_i$  and evolves freely thereafter, carrying the information about the system picked up at $t_i$
(this is an extension of von Neumann's model (vNM) of measurement
\cite{vNM,johansen_mello_2008}).
We may then detect that information at a later time;
we choose the detection time {\em for all} of the $N$ probes as $t_N^+$, i.e., right after the last coupling time $t_N$:
i.e., at $t_N^+$ we detect an observable of each one of the $N$ probes, not of the system itself; 
through the entanglement of system and probes we obtain information on the system observable $\hat{A}$ (see Fig. \ref{fig1}).

Another source of disturbance is $\hat{A}$ failing to commute with the total Hamiltonian $\hat{H}$ of the system and the probe: 
a system prepared in an eigenstate $|a_n\rangle$ of $\hat{A}$ would be found, in the course of time, in other eigenstates $|a_{n'}\rangle$ with non-zero probability.
This we remedy by requiring $[\hat{A},\hat{H}]=0$, so that $\hat{A}$ is a constant of motion, and thus a ``quantum non-demolition" (QND) observable \cite{imoto_et_al,caves_et_al,peres_book}.

From the detection at $t_N^+$ of the various {\em probes} successively coupled to {\em one} system we average the information acquired at $N$ successive times
and define what can aptly be called a {\em time average}. 
This average is performed on {\em one realization} $\omega$ of the QM ensemble.
We inquire whether, as $N\to \infty$, this result depends on the specific realization
and whether it coincides
--up to a set of zero measure-- 
with the expectation value of the detected 
observable for {\em one} probe performed over many 
realizations $\omega$ of the QM ensemble, 
which we shall call an {\em ensemble average}.
This is the question of {\em ergodicity} (see Fig. \ref{fig1}), as it is studied in the theory of stationary random processes \cite{yaglom}.
We then inquire whether from these averages we can find information on the 
system observable $\hat{A}$ as applicable to a QM ensemble.

The motivations of this paper are: 
i) to develop a theoretical analysis of an ergodic property
--relating time averages on a single system to QM ensemble averages--
which, to the best of our knowledge, has not been done in the past;
ii) to compare the theory with its actual materialization in true quantum non-demolition laboratory experiments.
The paper is organized as follows.
In Sec. \ref{vNM} we recall the von Neumann model of measurement for the system proper and $N$ probes,
and introduce the time-averaged operator $\overline{\hat{Q}}$ in 
Eq. (\ref{bar_hat_Q 1}) below.
The analysis of the question of ergodicity is performed in Sec. \ref{analysis ergodicity}, 
where we compute 
the QM average of $\overline{\hat{Q}}$ and its dispersion over the ensemble.
We then study how the reduced density operator of the system proper is affected by its interaction with the $N$ probes.
In Sec. \ref{decimation} we analyze the probability distribution (pd) of the eigenvalues of the observable $\hat{A}$ conditioned on the $N$ detected probe positions: we give an analytical treatment of the ``decimation" process that has been observed --and analyzed numerically-- in the experiments.
We finally conclude in Sec. \ref{summary}.

\section{The von Neumann model of measurement for the system proper and the $N$ probes.
The time-averaged operator.}
\label{vNM}

In the above {\em gedanken} experiment, an ``extended system" (ES) consisting of the system proper $s$ plus $N$ probes 
$\pi_1, \cdots \pi_N$ is considered.
Call $\omega$ {\em one preparation} of the ES, which we shall call {\em one realization}: 
the QM ensemble is the collection $\{ \omega \} \equiv \Omega$ of such realizations (see Fig. \ref{fig1}).

Each probe $\pi_i$ will be considered one-dimensional, with canonically conjugate dynamical variables $\hat{Q}_i,\hat{P}_i$ in the Schr\"odinger picture.
We define, for the ES, the operator
\begin{eqnarray}
\overline{\hat{Q}}
\equiv \frac1N \sum_{i=1}^{N}\hat{Q_i},
\label{bar_hat_Q 1}
\end{eqnarray}
which we call a ``time-averaged" operator, because the $N$ probes are applied at $N$ successive times.
The $\hat{Q_i}$'s commute among themselves:
we can thus detect the observable $\overline{\hat{Q}}$ by detecting, on the realization $\omega$ of the ensemble and at the fixed time $t_N^+$, the observables 
$\hat{Q_i}$'s (arising from probe $\pi_i$ each), and constructing the ``time-average"
\begin{equation}
\overline{Q^{(\omega)}_N}
= \frac1N \sum_{i=1}^{N}Q_i^{(\omega)} \; ,
\label{bar_Q 1}
\end{equation}
where $Q_i^{(\omega)}$ is the result of detecting $\hat{Q}_i$
in the realization $\omega$ (Fig. \ref{fig1}).
QM provides no way of calculating the time average $\overline{Q^{(\omega)}_N}$ of Eq. (\ref{bar_Q 1}), as it is the result of {\em one preparation}.
However, the standard rules of QM allow to compute the statistical properties of $\overline{Q^{(\omega)}_N}$ across a QM ensemble of preparations
(the set $\Omega$ introduced above).

We can now be more specific about the ergodicity question formulated above:
it will be answered by inquiring
whether the {\em time average} $\overline{Q^{(\omega)}_N}$ 
{\em over one realization} $\omega$ of a single ES depends on that realization, 
and whether, as $N\to \infty$, it coincides, up to a set of zero measure, with the 
expectation value
of one term in Eq. (\ref{bar_Q 1}), $Q_i^{(\omega)}$, over many 
realizations $\omega$ of the QM ensemble.
We remark that the model Hamiltonian used here --defined as the generalization of Eq. (\ref{V 2meas}) below to $N$ probes -- ensures {\em stationarity} 
in the sense of the theory of stationary random processes \cite{yaglom} 
[the expectation value in the final state of $\hat{Q}_i^m$ is independent of $i$; the correlation in the final state between 
$\hat{Q}_i$ and $\hat{Q}_j$ only depends on the difference $|i-j|$;
see comments below Eqs. (\ref{<Qi>f}), (\ref{<Qi2>f}), (\ref{<QiQj>f}), 
(\ref{ens average of t_av_1_syst 1 a}), (\ref{ens average of t_av_1_syst 1 b})].

\subsection{The case of two probes}
\label{N=2}

To illustrate the analysis, we consider the system proper $s$ coupled to only two probes $\pi_1$ and $\pi_2$, intended to ``pre-measure" the system observable $\hat{A}$ at $t=t_1$, and again at $t_2 \; (>t_1)$ with the same strength $\epsilon$.
Assume the two interactions to be of such a short duration that their time dependence can be approximated by delta functions at times $t_1$ and $t_2$, respectively.
We disregard the intrinsic dynamics of the system and of the probes and write the time-dependent Hamiltonian as \cite{caves,vNM,johansen_mello_2008,peres_book}
\begin{equation}
\hat{H} (t) = \epsilon \delta (t-t_1) \hat{A} \hat{P}_1
+ \epsilon \delta (t-t_2) \hat{A} \hat{P}_2 \; , \;\;\;\;
0 < t_1 < t_2 \; .
\label{V 2meas}
\end{equation}
The unitary evolution operator is given by \cite{johansen_mello_2008}
\begin{equation}
\hat{U}(t)
={\rm e}^{-\frac{i}{\hbar}\epsilon \theta(t-t_2) \hat{A} \hat{P}_2}
{\rm e}^{-\frac{i}{\hbar}\epsilon \theta(t-t_1) \hat{A} \hat{P}_1} \; ,
\label{U 2probes}
\end{equation}
where $\theta(t)$ is the Heaviside function. 
In the present model, with the evolution operator (\ref{U 2probes}),
$\hat{A}$ is a constant of motion and is thus a 
``non-demolition observable".
The model Hamiltonian (\ref{V 2meas}) can be generalized to interactions having a finite time duration, as long as they do not overlap in time 
(Ref. \cite{peres_book}, p. 350; Refs. \cite{kalev-mello,melloELAF}).

If the state of the system plus the two probes at $t=0$ is 
$
|\Psi \rangle_0 =  |\psi \rangle_s^{(0)}  |\chi \rangle_{\pi_1}^{(0)} 
|\chi \rangle_{\pi_2}^{(0)} \; ,
$
then for $t>t_2$, i.e., after the second interaction, it is given by
($f$ stands for ``final")
\begin{eqnarray}
|\Psi \rangle_f 
&=& {\rm e}^{-\frac{i}{\hbar}\epsilon  \hat{A} \hat{P}_2}
{\rm e}^{-\frac{i}{\hbar}\epsilon \hat{A} \hat{P}_1}
|\Psi \rangle_{0}
\nonumber \\
&=& \sum_n \left(\hat{\mathbb{P}}_{a_n} |\psi \rangle_s^{(0)}\right)
\left({\rm e}^{-\frac{i}{\hbar}\epsilon  a_n \hat{P}_1} 
|\chi \rangle_{\pi_1}^{(0)}\right)
\left({\rm e}^{-\frac{i}{\hbar}\epsilon  a_n \hat{P}_2} 
|\chi \rangle_{\pi_2}^{(0)}\right) \; .
\label{Psi_f}
\end{eqnarray}
The spectral representation 
$
\hat{A} = \sum_n a_n \hat{\mathbb{P}}_{a_n} 
$
was used, where $\hat{\mathbb{P}}_{a_n}$ denotes 
an eigenprojector of $\hat{A}$. 
The joint probability density (jpd) of the eigenvalues $Q_1, Q_2$ of the two probe-position operators for $t>t_2$ is
\begin{eqnarray}
&& p_f(Q_1,Q_2)
=  \;_f\langle \Psi | 
\hat{\mathbb{P}}_{Q_1} \hat{\mathbb{P}}_{Q_2}
|\Psi  \rangle _f
\nonumber \\
&=&  \sum_n
W_{a_n}^{(0)} \left|\chi_{\pi_1}^{(0)}(Q_1-\epsilon a_n) \right|^2 
\left|\chi_{\pi_2}^{(0)}(Q_2-\epsilon a_n) \right|^2 .
\label{p(Q1,Q2)} 
\end{eqnarray}
Here, 
$\hat{\mathbb{P}}_{Q_i}$ denotes 
an eigenprojector of $\hat{Q}_i$. 
The scalar product in (\ref{p(Q1,Q2)}) is understood to be evaluated with respect to all the degrees of freedom of the ES.
The quantity 
\begin{equation}
W_{a_n}^{(0)}
=\;  ^{(0)}_{\textcolor{white}{...}s} \langle \psi |\hat{\mathbb P}_{a_n}|\psi\rangle_s^{(0)}
\label{Wan}
\end{equation}
is the Born probability for the value $a_n$ in the original system state;
$\chi_{\pi_1}^{(0)}(Q_1-\epsilon a_n)$ is the shifted wave function of probe $\pi_1$ in the position representation, and similarly for $\pi_2$. 

\subsection{The arbitrary-$N$ case. The expectation value of probe positions.}
\label{N-arbitrary}

From Eq. (\ref{Psi_f}) generalized to $N$ probes, the jp amplitude for $a_n, Q_1, \cdots, Q_N$ is
\begin{equation}
\left\langle a_n, Q_1, \cdots, Q_N | \Psi \right\rangle_f
=\langle a_n|\psi_s^{(0)} \rangle \;
\prod_{i=1}^N\chi_{\pi_i}^{(0)}(Q_i-\epsilon a_n).
\label{Psi(an,Q1,...,QN)}
\end{equation}
For Gaussian packets with the same width $\sigma$ (the probe resolution) for the initial probe states \cite{caves,johansen_mello_2008}, the jpd's of $a_n, Q_1, \cdots, Q_N$ and of $Q_1, \cdots, Q_N$ are given by
\numparts
\begin{eqnarray}
p_f(a_n,Q_1,\cdots,Q_N)
= W_{a_n}^{(0)} \;
\prod_{i=1}^N \frac{{\rm e}^{-\frac{(Q_i-\epsilon a_n)^2}
{2\sigma^2}}}{\sqrt{2\pi \sigma^2}} \; , 
\label{pf(an,Q1,...,QN)} \\
p_f(Q_1,\cdots,Q_N)
=\sum_n 
 W_{a_n}^{(0)} \;
\prod_{i=1}^N \frac{{\rm e}^{-\frac{(Q_i-\epsilon a_n)^2}
{2\sigma^2}}}{\sqrt{2\pi \sigma^2}} \; .
\label{pf(Q1,...,QN)}
\end{eqnarray}
\label{pf(an,Q1,...,QN),pf(Q1,...,QN)}
\endnumparts
The Gaussian assumption allows an analytical treatment.
Use of Eqs. (\ref{pf(an,Q1,...,QN),pf(Q1,...,QN)})
and (\ref{pf(Q1,...,QN)}) gives
(the index $f$ indicates an expectation value evaluated with the state 
$|\Psi \rangle_f$)
\numparts
\begin{eqnarray}
\langle \hat{Q}_i \rangle_f 
&=& \langle \hat{Q}_1 \rangle_f 
= \sum_{n}W_{a_n}^{(0)} \cdot (\epsilon a_n) 
=\epsilon \langle \hat{A}\rangle_0 , \hspace{2cm}
\label{<Qi>f} \\
\langle \hat{Q}_i^2 \rangle_f 
&=& \langle \hat{Q}_1^2 \rangle_f 
=\sum_{n}W_{a_n}^{(0)} [(\epsilon a_n)^2 + \sigma^2]
=\epsilon^2 \langle \hat{A}^2\rangle_0 + \sigma^2.  \;\;\;\;\;\;
\label{<Qi2>f}
\end{eqnarray}
\label{<Qi>f,<Qi2>f}
\endnumparts
The notation $\langle \hat{A}\rangle_0$, $\langle \hat{A}^2\rangle_0$
indicates expectation values in the original system state.
The first equality in Eqs. (\ref{<Qi>f}), (\ref{<Qi2>f}) 
expresses the property of stationarity.
One can also show, for $i\neq j$
\begin{eqnarray}
\langle \hat{Q}_i \hat{Q}_j \rangle_f
= \langle \hat{Q}_1 \hat{Q}_2 \rangle_f  
= \sum_{n}W_{a_n}^{(0)} \cdot (\epsilon a_n) (\epsilon a_n) 
=\epsilon^2 \langle \hat{A}^2\rangle_0 \; .
\label{<QiQj>f}
\end{eqnarray}
In general, for a stationary random process,
$({\rm cov}(Q_i, Q_j))_f \equiv \langle \hat{Q}_i \hat{Q}_j \rangle_f 
- \langle \hat{Q}_i \rangle_f  \langle  \hat{Q}_j \rangle_f$
depends only on $|i-j|$.
In our present case,
$({\rm cov}(Q_i, Q_j))_f$
is independent of $i,j$ (for $i\neq j$) and does not decrease as $|i-j|$ increases.
This behavior is due to the structure of the jpd of $Q_1, \cdots, Q_N$ of 
Eq. (\ref{pf(Q1,...,QN)}).

\section{The analysis of the question of ergodicity}
\label{analysis ergodicity}

The {\em ensemble expectation value} over the realizations $\Omega$ of the time average $\overline{Q^{(\omega)}_N}$ of Eq. (\ref{bar_Q 1}) is given by
\numparts
\begin{eqnarray}
\left\langle \overline{Q^{(\omega)}_N} \right\rangle_{\Omega}
&=& \left\langle \overline{\hat{Q}}  \right\rangle_f
= \frac1N \sum_{i=1}^{N} 
\langle \hat{Q}_i  \rangle_f   
= \langle \hat{Q}_1 \rangle_f  
\label{ens average of t_av_1_syst 1 a}    \\
\left\langle \overline{\hat{Q}}  \right\rangle_f / \epsilon
&=& \langle \hat{A} \rangle_0 \; .
\label{ens average of t_av_1_syst 1 b}
\end{eqnarray}
\label{ens average of t_av_1_syst 1}
\endnumparts
We remark again that the same ensemble $\Omega$ of realizations is employed in the various QM expectation values appearing here and below (Fig. \ref{fig1}).
Eq. (\ref{ens average of t_av_1_syst 1 a}) 
states that the ensemble expectation value of the time average
$\overline{Q^{(\omega)}_N}$, which is just the QM expectation value of the
operator $\overline{\hat{Q}}$, 
coincides with the QM
expectation value of any one of the probe positions, like $\hat{Q}_1$.
This is the standard result in the theory of stationary random processes \cite{yaglom}.
Eq. (\ref{ens average of t_av_1_syst 1 b}) states that this result, in turn, equals, in units of $\epsilon$, $\langle \hat{A} \rangle_0$, 
the QM Born expectation value of the observable $\hat{A}$ in the original system state 
(the expectation value of $\hat{A}$ is time-independent, because 
$[\hat{A}, \hat{H}]=0$ in our model).

The crucial question is the dispersion of $\overline{Q^{(\omega)}_N}$ over the ensemble $\Omega$, which we calculate as
\numparts
\begin{eqnarray}
{\rm var} \overline{\hat{Q}}  
&=& 
\left\langle 
\left(\overline{\hat{Q}} 
- \left\langle \overline{\hat{Q}} \right\rangle_f \right)^2 
\right\rangle_f
= \left\langle 
\left(\overline{\hat{Q}} \right)^2 
\right\rangle_f
-\left\langle \overline{\hat{Q}} \right\rangle_f^2
\label{dispersion of t_av_1_syst 0}    \\ 
&=&  
\left\langle  \left( \frac1N \sum_{i=1}^{N}\hat{Q_i} \right)^2  \right\rangle_f  
-\left(\frac1N \sum_{i=1}^{N} \langle \hat{Q}_i  \rangle_f \right)^2       
\label{dispersion of t_av_1_syst 1}    \\                                     
&=& 
\frac{1}{N^2} 
\sum_{i,j = 1}^N
\left[
\langle \hat{Q}_i \hat{Q}_j \rangle_f
  -\langle \hat{Q}_i \rangle_f \langle \hat{Q}_j \rangle_f \right]
\label{dispersion of t_av_1_syst a} \\
&=& \frac{N-1}{N}({\rm cov} (\hat{Q}_1 , \hat{Q}_2))_f 
+\frac{1}{N}({\rm var}\hat {Q}_1)_f 
\label{dispersion of t_av_1_syst c} \\
& \to & ({\rm cov} (\hat{Q}_1 , \hat{Q}_2))_f , \;\; {\rm as} \; N \to \infty .
\label{dispersion of t_av_1_syst d}
\end{eqnarray}
Eq. (\ref{dispersion of t_av_1_syst 0}) is the familiar definition of the variance;
in Eq. (\ref{dispersion of t_av_1_syst 1}) we used the definition (\ref{bar_hat_Q 1})
to write the first and second moments of $\overline{\hat{Q}}$;
in Eq. (\ref{dispersion of t_av_1_syst c}) we used the first equality in 
Eqs. (\ref{<Qi>f}), (\ref{<Qi2>f}) and (\ref{<QiQj>f}).
Notice that ${\rm var} \overline{\hat{Q}}$ does not vanish in the limit $N\to \infty$.
This is not surprising, due to the behavior of the correlation function described right after Eq. (\ref{<QiQj>f}) \cite{yaglom}.
We can be more explicit using the last equalities appearing in 
Eqs. (\ref{<Qi>f}), (\ref{<Qi2>f}) and (\ref{<QiQj>f}), which give
\begin{eqnarray}
({\rm var} \overline{\hat{Q}})/\epsilon^2
&=& 
({\rm var}\hat{A})_0 + \frac{N_{cr}}{N}
\label{dispersion of t_av_1_syst 1 e} \\
& \ge & ({\rm var}\hat{A})_0 , \;\;\; {\rm if} \;\;\;\;\; N \gg N_{cr}\; ,
\label{dispersion of t_av_1_syst f}
\end{eqnarray}
\label{dispersion of t_av_1_syst}
\endnumparts
where we have defined a ``critical $N$",
$
N_{cr} = \left(\sigma / \epsilon \right)^2 \; .
$
If $N \ll N_{cr}$,
${\rm var} \overline{\hat{Q}}/ \epsilon^2 \gg ({\rm var}\hat{A})_0$;
this situation could be achieved with a very large probe resolution $\sigma$, measured in units of $\epsilon \sqrt{N}$.
If $N \gg N_{cr}$, 
${\rm var} \overline{\hat{Q}}/ \epsilon^2
\ge ({\rm var} \hat{A})_0$.
Given the resolution $\sigma$ of the probe, the coupling strength $\epsilon$  defines how quickly $N_{cr}$ is attained.

The conclusion is that the time average $\overline{Q^{(\omega)}_N}$, measured in units of $\epsilon$, has, in general, a {\em finite dispersion} over the ensemble $\Omega$ of realizations $\omega$, even when $N / N_{cr} \to \infty$.
Therefore, {\em we do not have ergodicity} and {\em we cannot verify the ensemble predictions of QM by means of a sequence of measurements on a single system}.
An exception is when the original QM system state is an eigenstate of the observable $\hat{A}$, 
as then
$({\rm var} \hat{A})_0=0$ 
and
${\rm var} \overline{\hat{Q}}/\epsilon^2 \to 0$ as $N \to \infty$.

\subsection{The probability distribution of $\overline{Q}_{N}$}
\label{p_f(Qbar-N)}

\begin{figure}[h]
\centerline{
\includegraphics[scale=0.6]{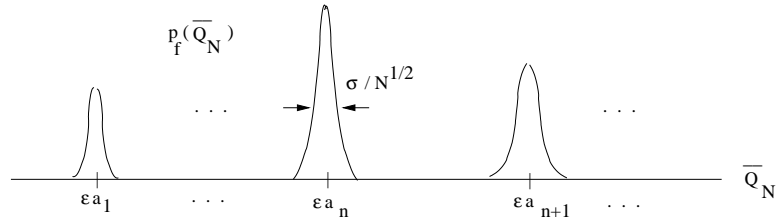}
}
\caption{
\footnotesize{Schematic illustration of the pd of $\overline{Q}_{N}$ of Eq. (\ref{p(barQN)}).
}
}
\label{fig2}
\end{figure}

From Eq. (\ref{pf(Q1,...,QN)}) we find the pd of $\overline{Q}_N$ of 
Eq. (\ref{bar_Q 1}) (sampled over the ensemble $\Omega$)
--whose first moment and variance were computed above-- as
\begin{equation}
p_f\left(\overline{Q}_{N}\right) 
= \sum_{n} W_{a_{n}}^{(0)} \;
\frac{{\rm e}^{-\frac{\left(\overline{Q}_{N}-\epsilon a_{n}\right)^2}
{2\frac{\sigma^2}{N}}}}
{\sqrt{2\pi\frac{\sigma^2}{N}}} \; ,
\label{p(barQN)}
\end{equation}
which is shown schematically in Fig. \ref{fig2}.
From Eq. (\ref{p(barQN)}) we see the effect of not having ergodicity.
The peaks in $p_f\left(\overline{Q}_{N} \right)$
are centered at the various $\epsilon a_n$'s. If $N\gg 1$, the peaks do not to overlap and the area under the peak centered at $\epsilon a_n$ gives $W_{a_{n}}^{(0)}$.
From one realization to another, $\overline{Q}_{N}$ jumps at random from one very narrow peak to another, $W_{a_n}^{(0)}$ being the fraction of realizations whose 
$\overline{Q}_{N} \approx \epsilon a_n$.
If the original system state is an eigenstate of $\hat{A}$, only one peak occurs and eventually we have ergodicity as $N \to \infty$. 

The process leading to one limiting value (as $N \to \infty$) of $\overline{Q}_{N}$ for a specific realization is most clearly seen in the extreme case 
$\sigma \ll \epsilon \Delta a_n$, when the original probe pd's are narrower than the separation of the $\epsilon a_n$'s.
We first detect $Q_1$:
from (\ref{pf(Q1,...,QN)}) for $N=1$, a result for $Q_1$ between two values 
$\epsilon a_n$ is extremely unlikely to occur.
If $Q_1 \approx \epsilon a_{n_0}$,
the jpd of $Q_2, \cdots, Q_N$, conditioned on 
$Q_1 = \epsilon a_{n_0}$, is, from Eq. (\ref{pf(Q1,...,QN)}), reduced approximately to one term
\begin{equation}
p_f(Q_2, \cdots, Q_N | Q_1 = \epsilon a_{n_0}) \approx 
\prod_{i=2}^{N} \frac{{\rm e}^{-\frac{\left(Q_i-Q_1\right)^2}
{2\sigma^2}}} 
{\sqrt{2\pi\sigma^2}} \; .
\label{p(Q2...QN|Q1)}
\end{equation}
Having found $Q_1= \epsilon a_{n_0}$, it is as if $Q_2, \cdots, Q_N$ were statistically independent variables, their pd's all centered at $\epsilon a_{n_0}$ and with a width $\sigma$; i.e., the first detected value $Q_1$ makes $Q_2, \cdots, Q_N$  
get ``stuck" around $\epsilon a_{n_0}$. 
As a result, $\bar{Q}_N$ tends to the limiting value $\epsilon a_{n_0}$ as 
$N\to \infty$.
Had we found $Q_1= \epsilon a_{n_1}$, $Q_2, \cdots, Q_N$ would be stuck around 
$\epsilon a_{n_1}$ and $\bar{Q}_N$ would tend to the limiting value 
$\epsilon a_{n_1}$.
As a matter of fact, the probability distribution of $\bar{Q}_N$, conditioned on 
$Q_1 = \epsilon a_{n_0}$, is found to be
\begin{equation}
p_f(\bar{Q}_N| Q_1 = \epsilon a_{n_0})
= \frac{{\rm e}^{-\frac{\left(\bar{Q}_N-Q_1\right)^2}
{2\frac{N-1}{N^2}\sigma^2}}} 
{\sqrt{2\pi \frac{N-1}{N^2}  \sigma^2}} \; .
\label{p(QbarN|Q1)}
\end{equation}
A corresponding analysis can be carried out in the opposite extreme case 
$\sigma \gg \epsilon \Delta a_n$, when the original probe pd's are wider than the separation of the $a_n$'s.

\subsection{The reduced density operator of the system proper}
\label{rho-s-f}
To clarify to what extent has the system proper $s$ been altered due to its interaction with the $N$ probes, we calculate the final reduced density operator of the system.
Tracing over $\pi_1, \cdots \pi_N$ the density operator 
${|\Psi\rangle_{f}} {_{f}\langle \Psi|}$ from
Eq. (\ref{Psi_f}) generalized to $N$ Gaussian probes, we find
\begin{eqnarray}
\rho_s^{(f)}
= \sum_{nn'}
e^{-\frac{N}{8 N_{cr}} (a_n - a_{n'})^2}
\Big(
\mathbb{P}_{a_n} |\psi \rangle_s^{(0)} \; ^{(0)}_{\textcolor{white}{..}s}\langle \psi | \mathbb{P}_{a_{n'}}
\Big) \; .
\label{rho-sf N}
\end{eqnarray}
The non-demolition property is clear:
the diagonal matrix elements of $\rho_s^{(f)}$ are unchanged by the interaction with the probes; the off-diagonal ones are changed, depending on the interaction strength $\epsilon$.
For $N$ probes and $N \ll N_{cr}$, 
\numparts
\begin{eqnarray}
\rho_s^{(f)} 
\approx |\psi \rangle_s^{(0)} \; ^{(0)}_{\textcolor{white}{..}s} 
\langle \psi | \; ,
\label{rho-sf N 1}
\end{eqnarray}
and the system state is not altered appreciably by the detections. 
For $N \gg N_{cr}$, 
\begin{eqnarray}
\rho_s^{(f)}
\approx
\sum_n \mathbb{P}_{a_n} |\psi \rangle_s^{(0)} \; 
_{\textcolor{white}{...}s}^{(0)}\langle \psi | \mathbb{P}_{a_{n}} \; , 
\label{rho-sf N 2}
\end{eqnarray}
\endnumparts
and the final state is a mixture like the one found after a non-selective projective measurement on the original pure state \cite{johansen07}, 
a result eventually attained as $N$ increases, no matter how small
--but fixed-- is $\epsilon/\sigma$.
However, the final system state can be kept close to the original one for $N$ as large as we please, if $N_{cr}$ is large enough.

The $N$ dependence of the transition between
Eqs. (\ref{rho-sf N 1}) and (\ref{rho-sf N 2}) exhibits the progressive modification of the density operator for the system proper resulting from the process.    
We complement this discussion in the next section, where we study the mechanism behind what has been called
by the experimental groups \cite{guerlin_et_al07,bauer-bernard} the ``progressive collapse" of the system state.



\section{Probability distribution of the $a_n$'s, conditioned on the detected values for the probes.}
\label{decimation}

Refs. \cite{brune,guerlin_et_al07,bauer-bernard}
analyze the change suffered by the pd of the photon number in the cavity, conditioned on the detection of the $N$ probes.
Here we use our model, in which the probes interact with the system at $N$ successive times and the detection of the $N$ probes takes place at the single time $t_N^+$.
Starting from Eq. (\ref{pf(an,Q1,...,QN)}): i) integration over 
$Q_1, \cdots, Q_{N}$ gives
$
W^{(f)}_{a_n}=W^{(0)}_{a_n} \; ,
$
as a consequence of the non-demolition character of the vNM Hamiltonian [see below Eq. (\ref{rho-sf N})];
ii) the $a_n$ pd 
conditioned on the detected $Q_1, \cdots, Q_{N}$ 
for each preparation of the ES is
\begin{eqnarray}
&& p_f\left(a_n|Q_1,\cdots,Q_{N}\right) 
\nonumber \\
&& \hspace{8mm} = W_{a_n}^{(0)} 
\frac
{{\rm e}^{-\frac{\left(a_n - \overline{Q}_{N}/\epsilon\right)^2}
{\frac2N \left(\frac{\sigma}{\epsilon}\right)^2}}}
{\sum_{n'} W_{a_{n'}}^{(0)} \; 
{\rm e}^{-\frac{\left(a_{n'} - 
\overline{Q}_{N}/\epsilon\right)^2}
{\frac2N \left(\frac{\sigma}{\epsilon}\right)^2}} }  \; .
\label{pf(n|Q1,...,QN)} 
\end{eqnarray}
Gaussian probe functions make 
$p_f\left(a_n|Q_1,\cdots,Q_{N}\right)$ 
depend on the probe positions only through $\overline{Q}_{N}$.

The pd of the system $a_n$'s, which is originally $W_{a_n}^{(0)}$, after its interaction with the probes and conditioned on a 
specific $N$-tuple $Q_1, \cdots Q_N$ of probe positions, has become modulated by the second factor in Eq. (\ref{pf(n|Q1,...,QN)}), 
which ``disects" it into a slice centered at 
$a_n \sim \overline{Q}_{N}/\epsilon$.
From the  above discussion, Eq. (\ref{p(barQN)}) to 
Eq. (\ref{p(QbarN|Q1)}), the centroid of the disecting factor eventually tends to a limiting value as $N$ increases;
at the same time, its width,
$\sigma / (\epsilon \sqrt{N})$, becomes thinner the larger is $N$.
This is the ``decimation process" of Ref. \cite{brune}, 
Fig. 2 of Ref. \cite{guerlin_et_al07} and Ref. \cite{bauer-bernard},
where probe functions arise from a Ramsey-interferometer-type experimental setup and decimation is exhibited numerically; 
the present model allows an analytical treatment.
We think it is of interest to have pointed out explicitly the above mechanism applied to the experiments we have been referring to, because, to the best of our knowledge, it has not been put in the language of ergodicity.

Eq. (\ref{pf(n|Q1,...,QN)}) gives the pd of $a_n$ 
conditioned on a given set of probe positions 
$Q_1, \cdots, Q_N$;
the disecting factor is centered at 
$a_n$ $\sim \overline{Q}_N/\epsilon$.
For a different value of $\overline{Q}_N$, say $\overline{Q}'_{N}$, 
it is centered at 
$\overline{Q}'_{N}/\epsilon$.
Running through the ensemble, the disections in 
Eq. (\ref{pf(n|Q1,...,QN)}) appear with the frequency of occurrence of $\overline{Q}_{N}/\epsilon$, i.e. $p_f\left(\overline{Q}_{N}/\epsilon\right)$, obtainable from Eq. (\ref{p(barQN)}).
Accumulating all the $\overline{Q}_{N}/\epsilon$, i.e. constructing
$\int p_f(a_n|\overline{Q}_{N}/\epsilon)p_f\left(\overline{Q}_{N}/\epsilon\right)
d \overline{Q}_{N}/\epsilon$,
we recover the original pd $W_{a_{n}}^{(0)}$ of the $a_n$'s, just as observed in 
Ref. \cite{guerlin_et_al07}, Fig. 3.

\section{Summary}
\label{summary}

In summary, we investigated whether ergodicity is realized in QM. 
To control the disturbance produced by the measurement,
we required the observable $\hat{A}$ to be a non-demolition one, and
we introduced $N$ probes which interact with the system at successive times $t_i$ with a coupling strength $\epsilon$, and we detect the probes. 
This scheme has been materialized in QED-cavity experiments, where the probes are atoms that traverse the cavity at successive times and are then detected.
In general, the system is {\em not ergodic}:
thus, from the time average over {\em one} realization of the system plus $N$ probes, we cannot infer the QM ensemble average.

The reduced density operator for the system is not appreciably altered by the $N$ detections if $N \ll N_{cr}$.
If $N \gg N_{cr}$, an initially pure state eventually becomes a complete mixture.

The probability of the eigenvalues $a_n$, conditioned on the detected positions of the $N$ probes, is the original Born probability modulated by a factor that depends on $\overline{Q}_N$ for the detected values (decimation process).
Probe Gaussian functions allow an analytical treatment.

The statistical distribution over the QM ensemble of  $\overline{Q}_N$ consists of a series of peaks centered at the eigenvalues $\epsilon a_n$:
the presence of more than one peak is a consequence of not having ergodicity.

Finally, we wish to comment that the analysis carried out in this paper is fully quantum mechanical.
It is, however, interesting to mention that the classical counterpart of our Hamiltonian, 
Eq. (\ref{V 2meas}), (see, e.g., Ref. \cite{peres_book}, pp. 378-380), gives a joint probability distribution of the probe positions $p_f(Q_1,Q_2)$ with a similar structure of our 
Eq. (\ref{p(Q1,Q2)}) for $N=2$ probes, or its generalization (\ref{pf(Q1,...,QN)}) for an arbitrary number of probes.
The complementary part of our study, the density operator $\rho_s^{(f)}$ for the system proper, has clearly a fully QM structure, as can be seen from Eq. (\ref{rho-sf N}).


\ack
One of the authors (PAM) acknowledges financial support by DGAPA, Mexico, under Grant IN109014.



\end{document}